\documentclass{appolb}
\usepackage{epsfig}
\usepackage{amsmath}
\usepackage{subfigure}
\begin{document}
\bibliographystyle{h-elsevier3} 
\title{Forward-backward rapidity correlations in relativistic heavy ion collisions : torqued fireball%
\thanks{Presented at \textit{Strangeness in Quark Matter 2011}, Sept. 18--24, Cracow, Poland}%
}
\author{
Joao Moreira \address{Centro de F\'{i}sica Computacional, Department of Physics,
University of Coimbra, 3004-516 Coimbra, Portugal}
\and
Piotr Bo\.zek
\address{The H. Niewodnicza\'{n}ski Institute of Nuclear Physics, Polish Academy of Sciences, PL-31342 Krak\'{o}w, Poland\\
Institute of Physics, Rzesz\'{o}w University, PL-35959 Rzesz\'{o}w, Poland}
\and
Wojciech Broniowski
\address{The H. Niewodnicza\'{n}ski Institute of Nuclear Physics, Polish Academy of Sciences, PL-31342 Krak\'{o}w, Poland\\
Institute of Physics, Jan Kochanowski University, PL-25406 Kielce, Poland}
}
\maketitle
\begin{abstract}
Statistical fluctuations in the transverse distribution of sources in relativistic heavy ion collisions and an asymmetric emission profile associated with the wounded nucleons lead to rapidity dependence of the reaction plane. The size of this effect is estimated for the gold-gold collisions at the highest RHIC energy ($\sqrt{s_{NN}}=200~\mathrm{GeV}$) in the Glauber model using {\tt GLISSANDO}. The hydrodynamical evolution of the resulting torqued fireball is carried out in $3+1$ dimensions considering a perfect fluid. Hadronization is simulated using {\tt THERMINATOR} including non-flow contribution coming from resonance decays. Some experimental measures that can be used to detect the torque effect are proposed.
\end{abstract}
\section{Introduction}
\begin{figure}[tb]
\label{torqueorigin}
\begin{center}
\subfigure[]{\label{torqueorigina}\includegraphics[width=0.25\textwidth]{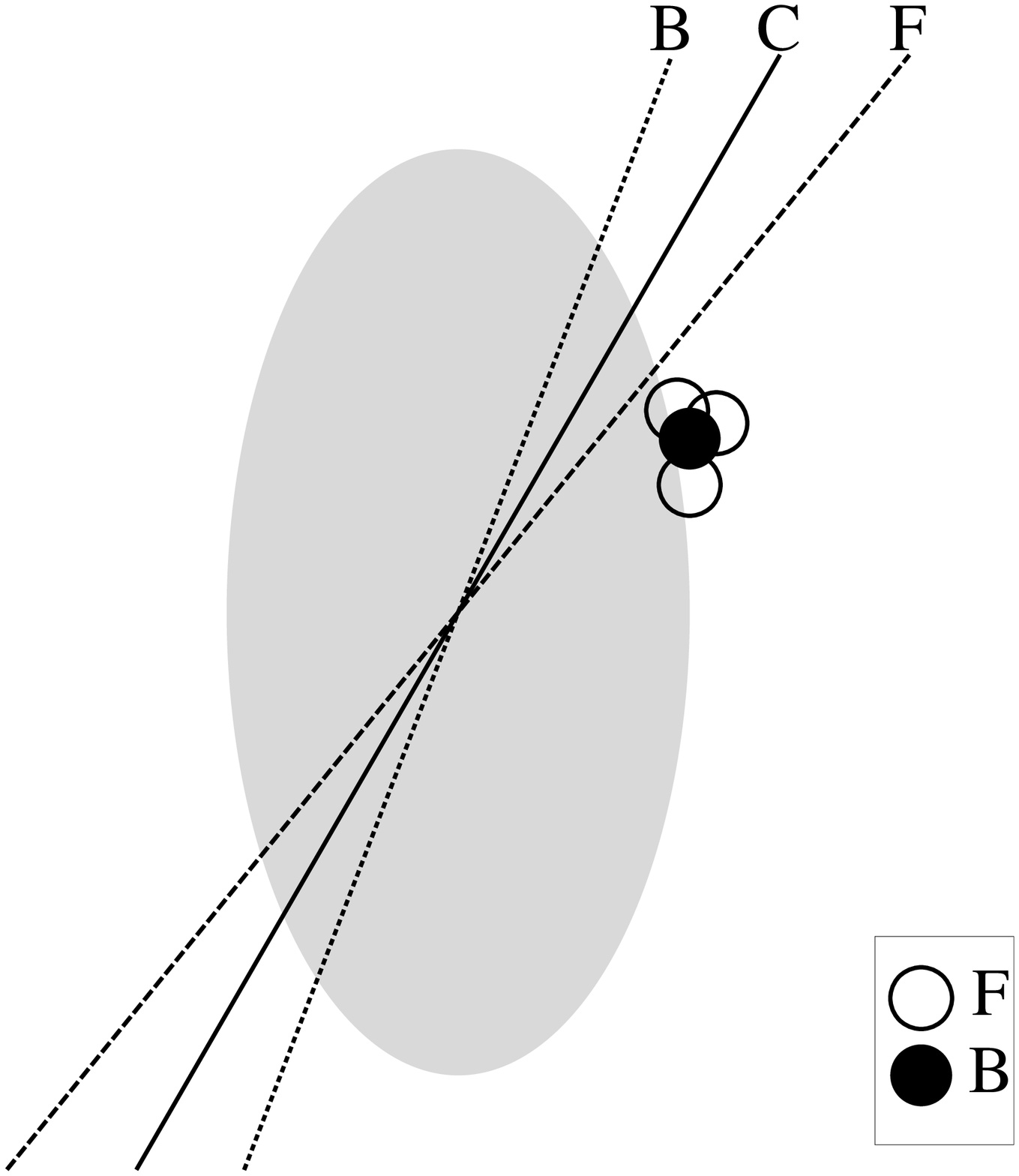}}
\subfigure[]{\label{torqueoriginb}\includegraphics[width=0.38\textwidth]{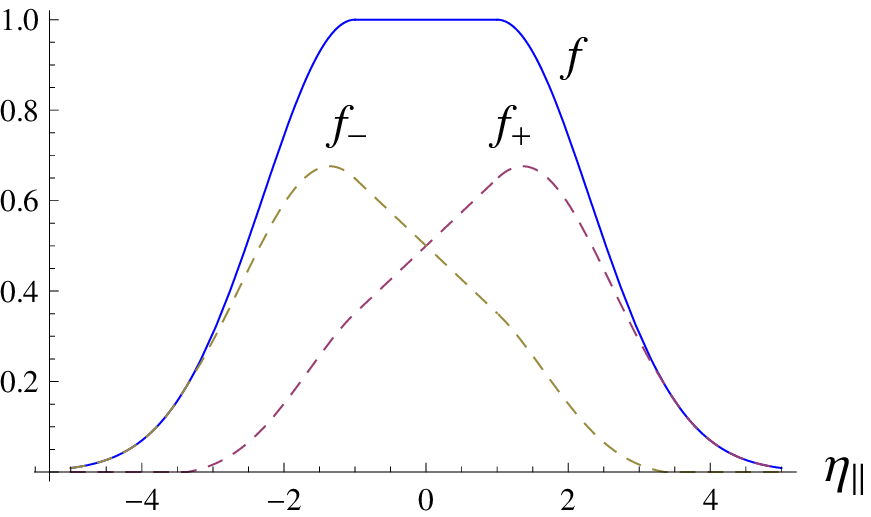}}
\subfigure[]{\label{torqueoriginc}\includegraphics[width=0.25\textwidth]{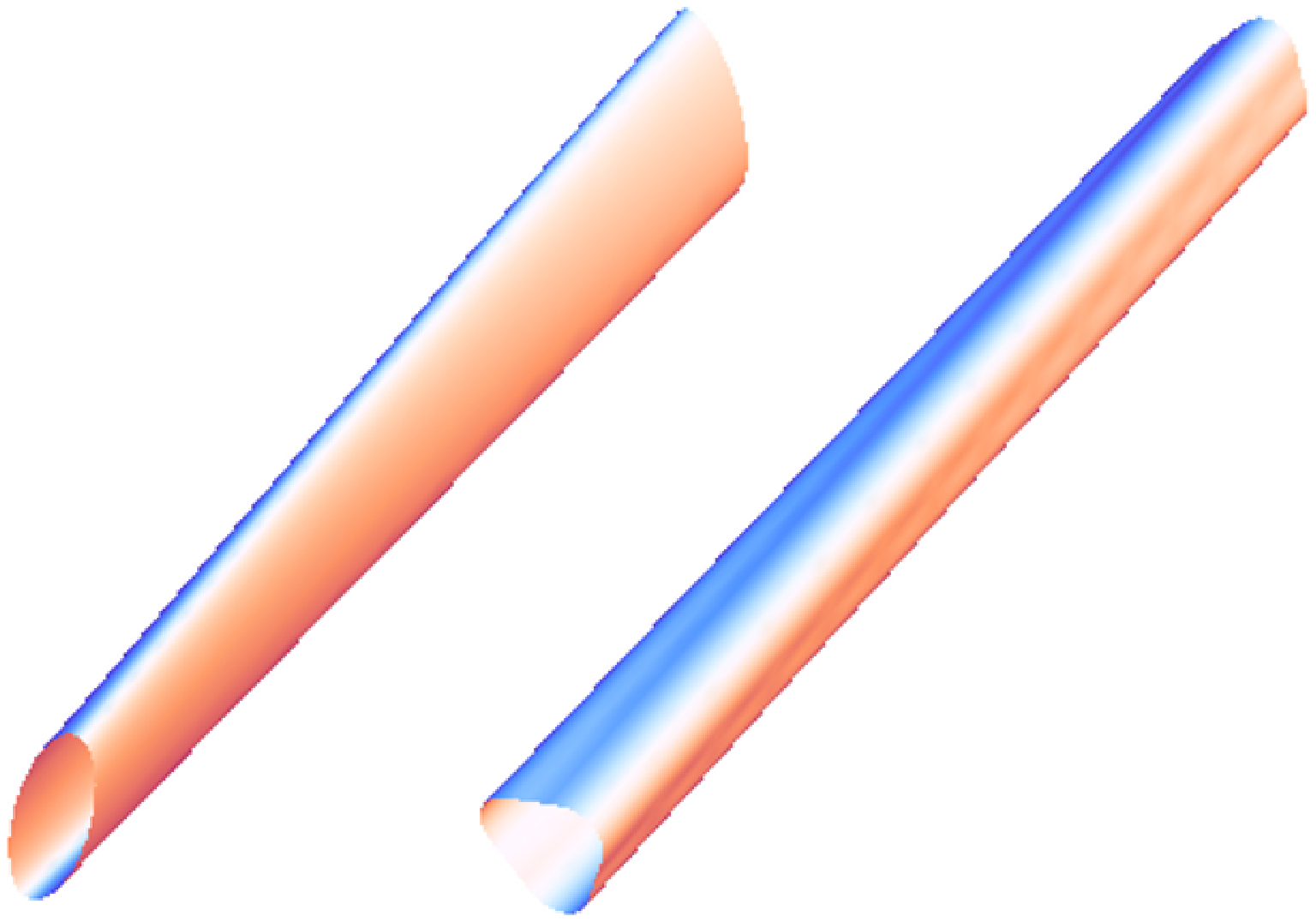}}
\caption{(Color online) The inclusion of a random asymmetric cluster, schematically depicted in Fig. \ref{torqueorigina}, twists the principal axis. Due to the asymmetry in the emission profiles of forward(backward) moving wounded nucleons, $f_\pm$, seen in Fig. \ref{torqueoriginb}, this twist is rapidity dependent ($f$ refers to the emission profile for binary collisions). The schematic figure of the torqued fireball, Fig. \ref{torqueoriginc}, elongated along the $\eta_\parallel$ axis.  The direction of the principal axes in the transverse plane  rotates as $\eta_\parallel$ increases. The left and right pictures correspond to the rank-2 (elliptic) and rank-3 (triangular) cases, respectively. The effect occurs event-by-event.}
\end{center}
\end{figure}

Dynamical properties of the early stage of relativistic heavy ion collisions can be probed using long range rapidity correlations. Starting from the wounded nucleon \cite{Bialas:1976ed,Bialas:2008zza} approach to the initial stage of heavy ion collisions a torqued fireball picture naturally arises as a result \cite{Bozek:2010vz}: the principal axes are twisted in opposite directions in the forward and backward regions. The underlying reasons for this are the statistical fluctuations in the transverse density of the sources (wounded nucleons) and the assymetry of the particle emission function (forward (backward) moving wounded nucleons emit particles preferably in the forward (backward) direction). In Fig. \ref{torqueorigina} we represent schematically a cluster of wounded nucleons causing a twist in the principal axes. If we are dealing with a different number of forward and backward moving wounded nucleons, due to the asymmetry of the emission functions, this twist is rapidity dependent. The initial energy density is:
\begin{small}
\begin{align}
F(\eta_\parallel,x,y)&=(1-\alpha)[\rho_+(x,y) f_+(\eta_\parallel)+ \rho_-(x,y) f_-(\eta_\parallel)]+\alpha 
\rho_{\rm bin}(x,y) f(\eta_\parallel), 
\label{eq:em}
\end{align}
\end{small}
where $\rho_\pm$, $\rho_{\rm bin}$ are the density of forward (backward) moving wounded nucleons and binary sources. The respective emission profiles $f_\pm$, $f$ are depicted in Fig. \ref{torqueoriginb} (in the chosen parametrization $f=f_++f_-$ and is symmetric in rapidity). The parameter $\alpha$ corresponds to the fraction of binary sources in this mixed model \cite{Kharzeev:2000ph,Back:2001xy,Back:2004dy}.

The detection of this effect from the data consisting of the final momenta of observed particles in an event is a challenge as one still has to take into account the effect of the fireball expansion and hadronization.
\section{Simulation results}
\begin{figure}[htp]
\begin{center}
\subfigure[]{\label{FBdiff2}\includegraphics[width=0.43\textwidth]{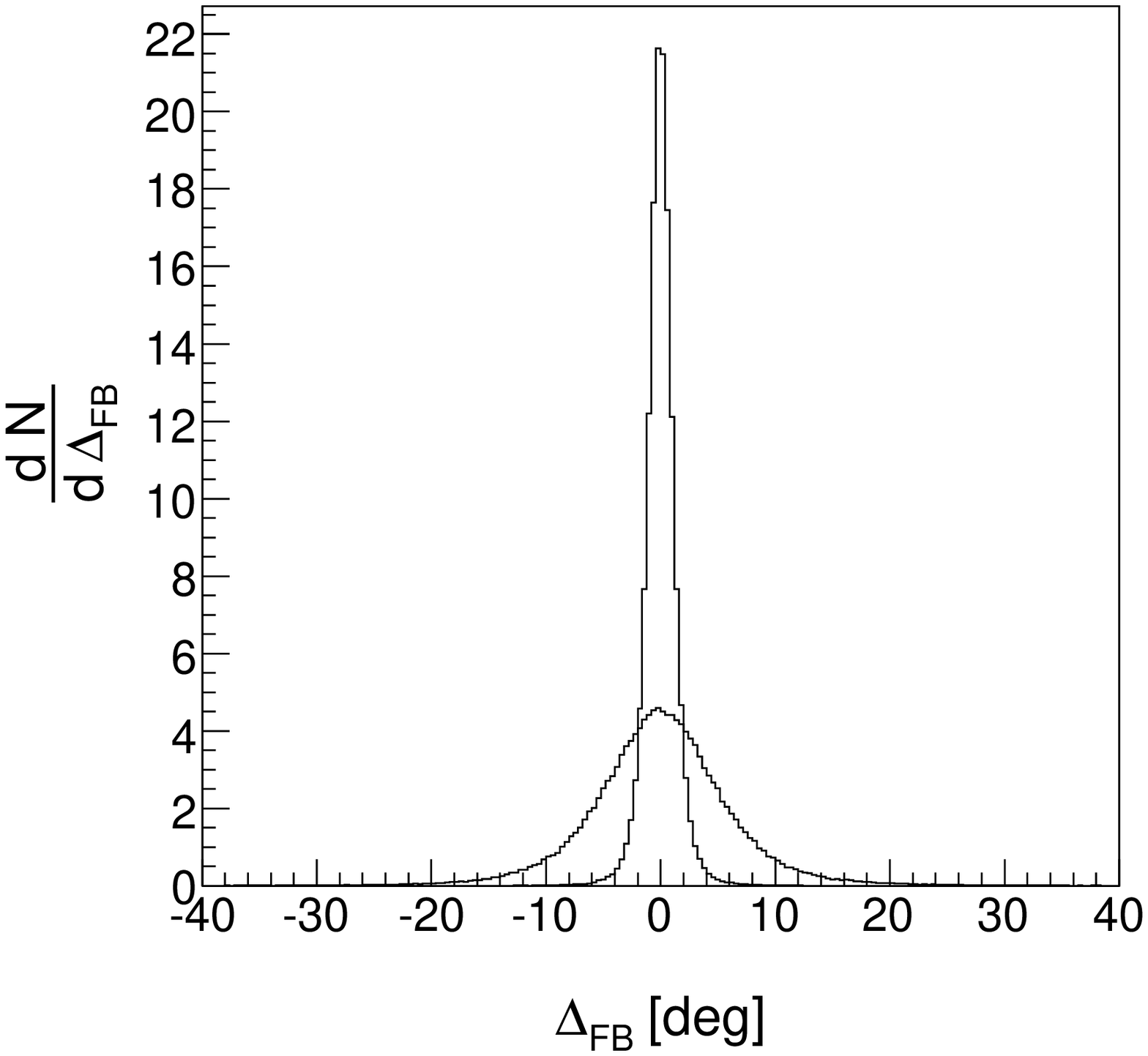}}
\subfigure[]{\label{FBdiff2sigma}\includegraphics[width=0.43\textwidth]{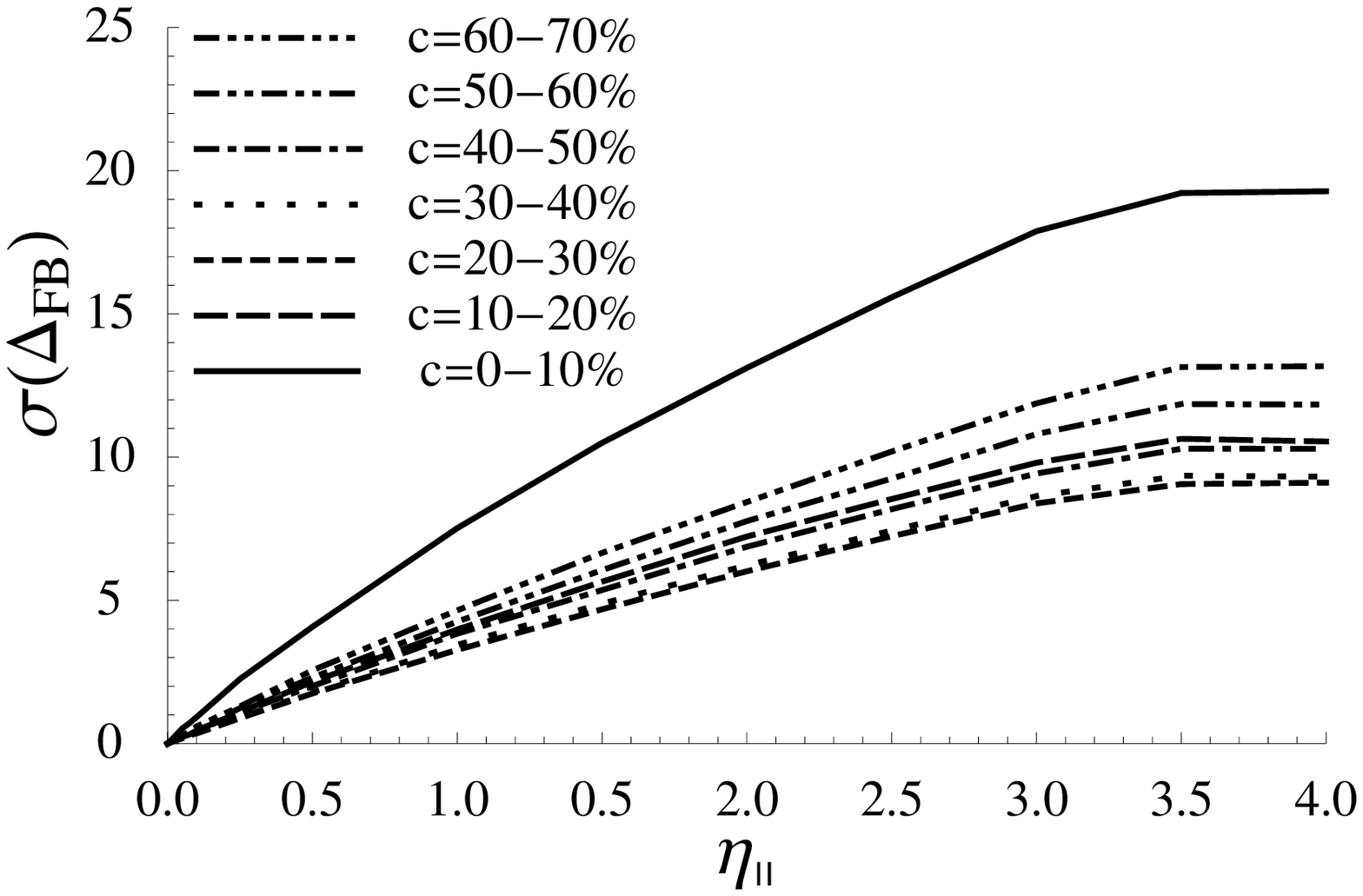}}
%\subfigure[]{\includegraphics[width=0.48\textwidth]{Imagens/distr2Dphi2FBb73eta25}}
%\subfigure[]{\includegraphics[width=0.48\textwidth]{Imagens/CovDFBCphi2vsrap}}
%\subfigure[]{\includegraphics[width=0.24\textwidth]{Imagens/diffangles3overlayb73}}
%\subfigure[]{\includegraphics[width=0.24\textwidth]{Imagens/sigmaDFBphi3vsrap}}
%\subfigure[]{\includegraphics[width=0.24\textwidth]{Imagens/distr2Dphi3FBb73eta25}}
%\subfigure[]{\includegraphics[width=0.24\textwidth]{Imagens/CovDFBCphi3vsrap}}
\end{center}
\caption{Distribution of the forward-backward angle difference for the elliptic deformation, $\Delta_{FB}$, for the $20-30\%$ centrality class in Fig. \ref{FBdiff2} for $\eta_\parallel=0.5$ and $\eta_\parallel=2.5$ (the latter is the wider distribution). Rapidity dependency of the rms of $\Delta_{FB}$ for several centrality classes Fig. \ref{FBdiff2sigma}.}
\end{figure}

To evaluate the size of the torque effect in the initial fireball we perform a Monte-Carlo simulation using {\tt GLISSANDO} \cite{Broniowski:2007nz} in the Glauber mixed model.  In a given event, the angle of the principal axis for the Fourier moment of rank-$k$ for $n$ sources at some $\eta_\parallel$ is given by:
\begin{small}
\begin{align}
\Psi^{(k)}&=\frac{1}{k} {\rm arctan} \left ( \frac{\sum_{i=1}^{n} w_i r_i^2 \sin(k \phi_i)}{\sum_{i=1}^{n} w_i r_i^2 \cos(k \phi_i)} \right ).
\end{align}
\end{small}
Here the weight of the sources are: $w_i=(1-\alpha)f_{\pm}$ for forward(backward) moving wounded nucleons and $w_i=\alpha f$ for binary. We then define the differences: $\Delta_{FB}=\Psi^{(2)}(\eta_\parallel)-\Psi^{(2)}(-\eta_\parallel)$ and $\Delta_{FB(BC)}=\Psi^{(2)}(\pm\eta_\parallel)-\Psi^{(2)}(0)$. The magnitude and sign of the resulting torque angle fluctuate from event to event. For Au+Au collisions at the highest available energy at RHIC the expected size of $\Delta_{FB}$ is of $20^\circ$ for central and $10^\circ$ for mid-peripheral collisions (see Fig. \ref{FBdiff2sigma}). 

\begin{figure}[htp]
\begin{center}
\label{Distr2D}
\subfigure[]{\label{HydroEvo}\includegraphics[width= 0.36 \columnwidth]{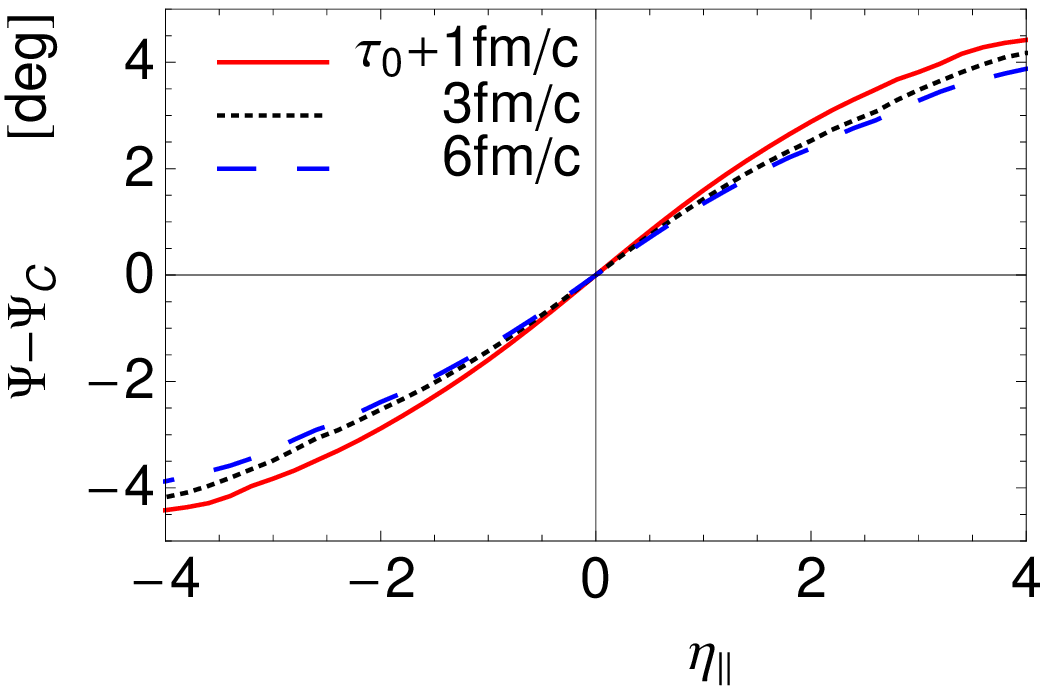}}
\subfigure[]{\label{anglespread}\includegraphics[width=0.36\textwidth]{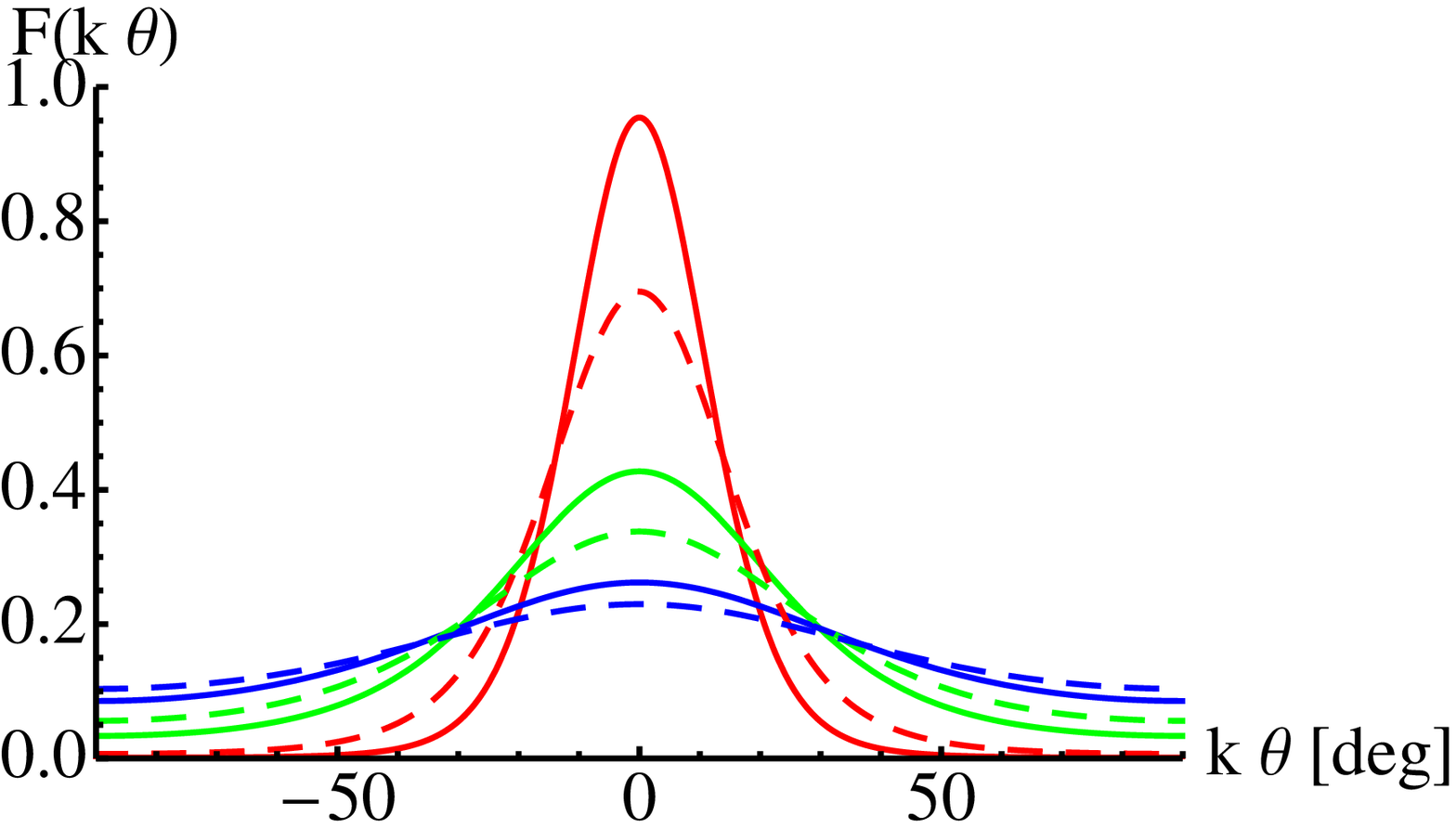}}
\caption{Angle as a function of space-time rapidity for different values of proper time given by 3+1 perfect hydrodynamic evolution of the fireball can be seen in Fig. \ref{HydroEvo}. Estimate of the difference between the fireball angle and the event angle for three different multiplicities and two different values of the flow coefficient (red, green and blue correspond respectively to $n=600,~100$ and $20$; solid lines correspond to $v_k=5\%$ and dashed to $2\%$).}
\end{center}
\end{figure}

The $3+1$ dimensional (perfect fluid) hydrodynamical simulation of the fireball expansion was done as in \cite{Bozek:2009ty}. The results suggest that the effect survives this phase with almost no change in its size, as can be seen in Fig. \ref{HydroEvo}. The energy density used as a starting point for the hydrodynamical simulation is given by:
\begin{small}
\begin{align}
\epsilon(\eta_\parallel,x,y)=&(1-\alpha)[\rho_+(Rx,Ry) f_+(\eta_\parallel) + \rho_-(R^Tx,R^Ty) f_-(\eta_\parallel)] +\alpha \rho_{\rm bin}(x,y) f(\eta_\parallel),
\end{align}
\end{small}
where the operators $R$ and $R^T$ rotate the density of the forward and backward moving wounded nucleons (respectively) by a fixed angle in opposite directions ($5^\circ$ for the centrality $c=20-25\%$).

However, this effect is significantly washed away in the hadronization stage. Using the result from  Appendix C of \cite{Broniowski:2007ft} we can estimate the deviation of the event angle from the fireball angle, as seen in Fig. \ref{anglespread}. Higher multiplicity as well as higher flow coefficient lower the statistical noise.
\section{Proposed measures: cumulants}
\begin{figure}[htp]
\begin{center}
\label{cumulants}
\subfigure[]{\label{prim_c2}  \includegraphics[width= 0.36\columnwidth]{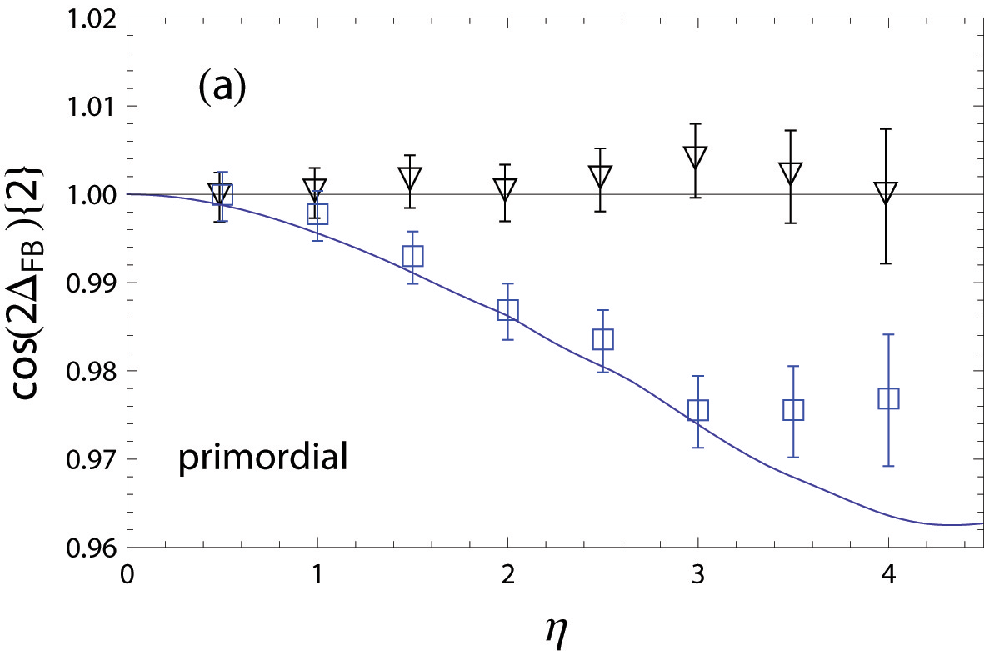}}
\subfigure[]{\label{above045a}\includegraphics[width= 0.36\columnwidth]{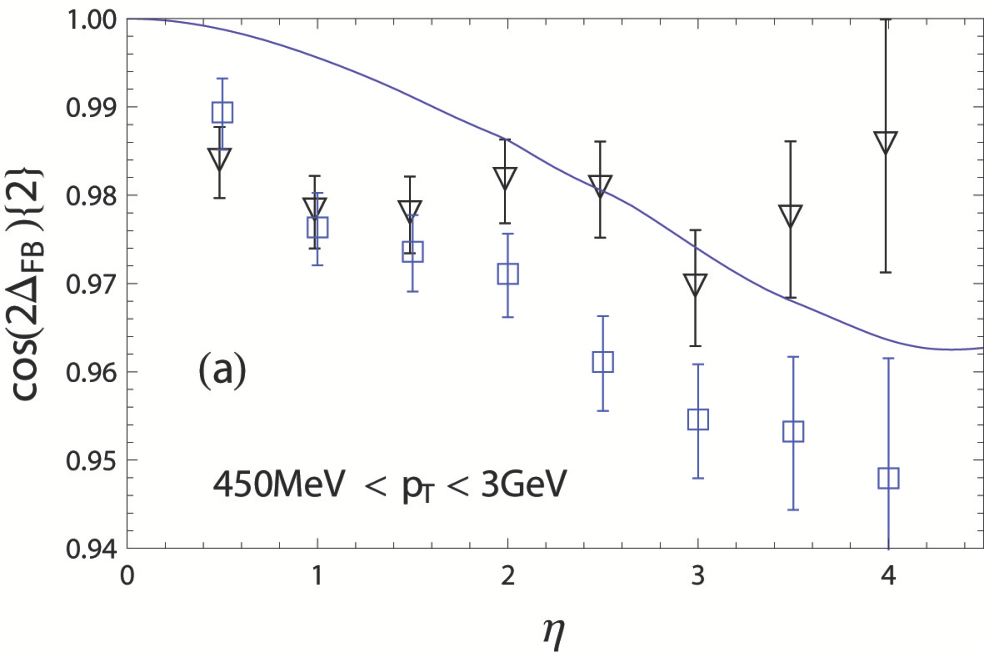}}
\subfigure[]{\label{prim_c4}  \includegraphics[width= 0.36\columnwidth]{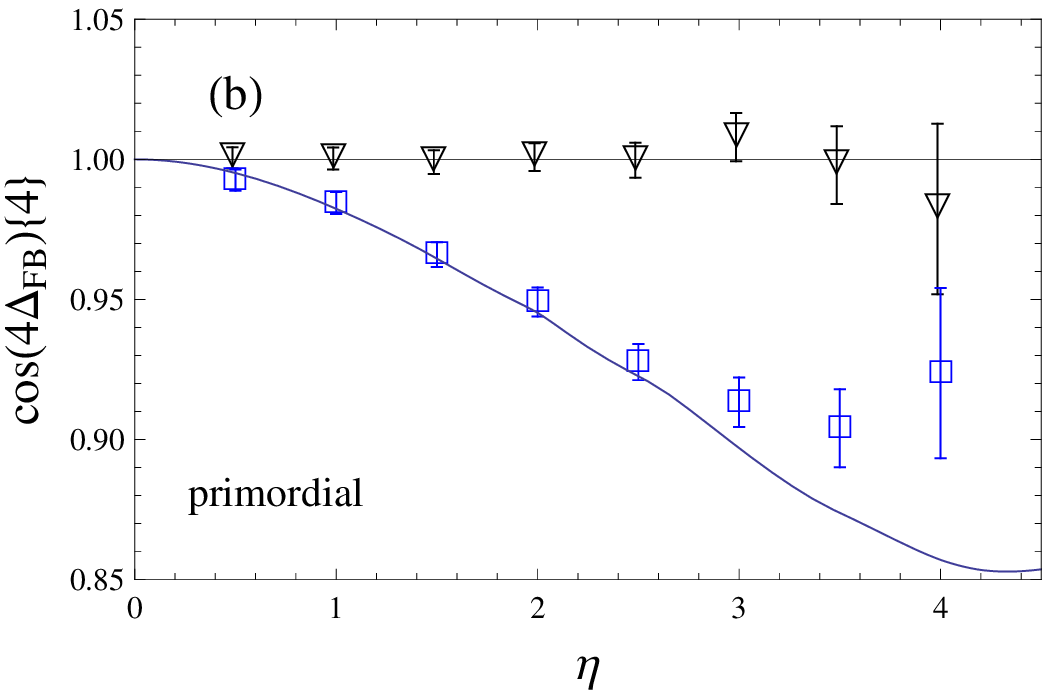}}
\subfigure[]{\label{above045b}\includegraphics[width= 0.36\columnwidth]{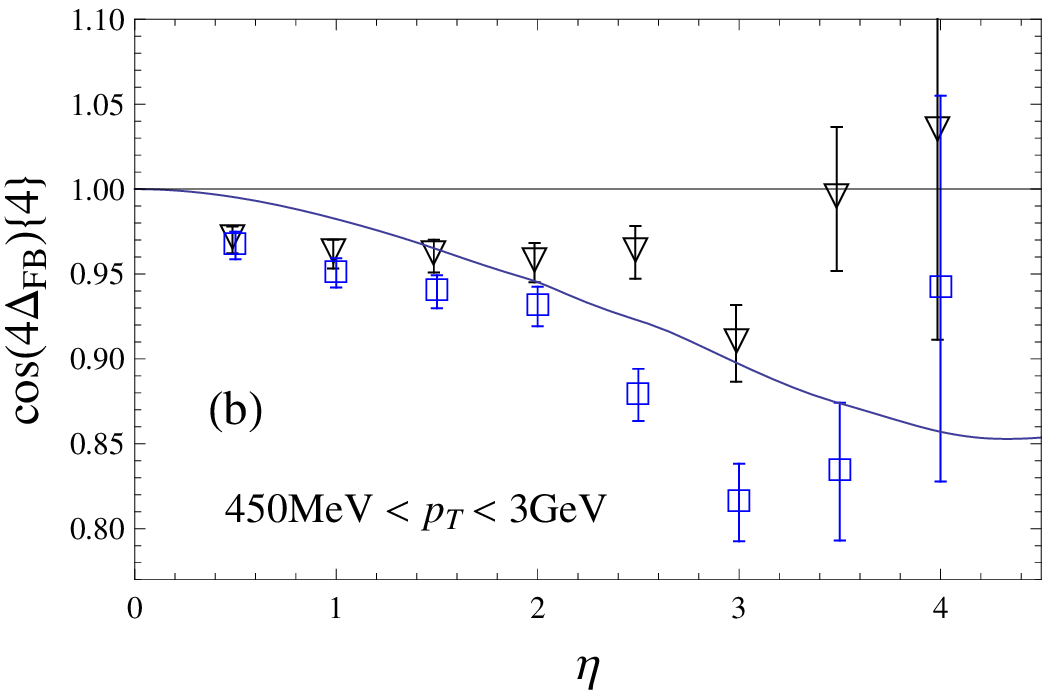}}
\subfigure[]{\label{AFBC_prim}\includegraphics[width= 0.36\columnwidth]{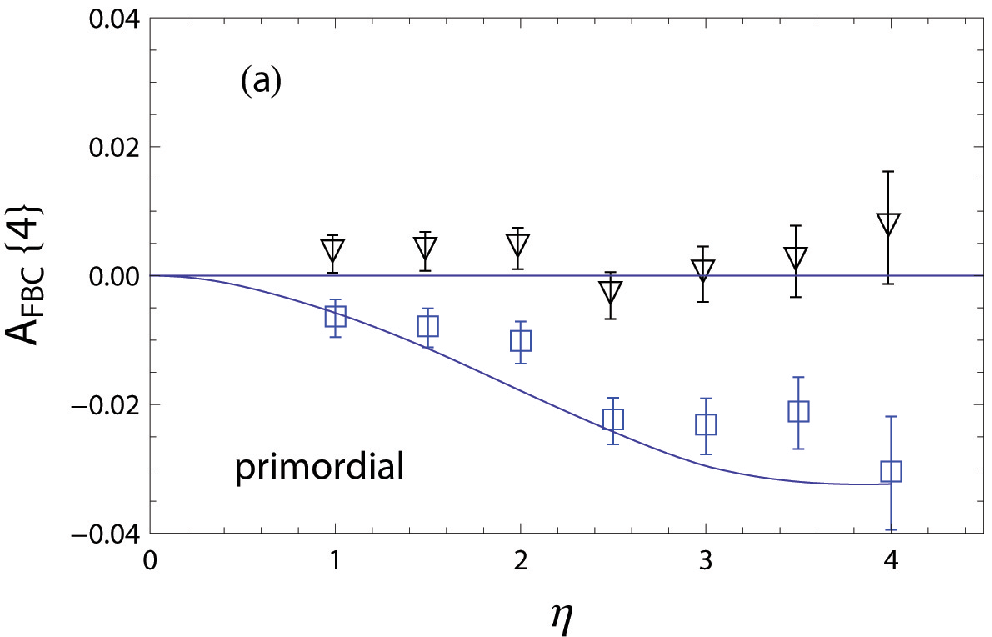}}
\subfigure[]{\label{AFBC}\includegraphics[width= 0.36\columnwidth]{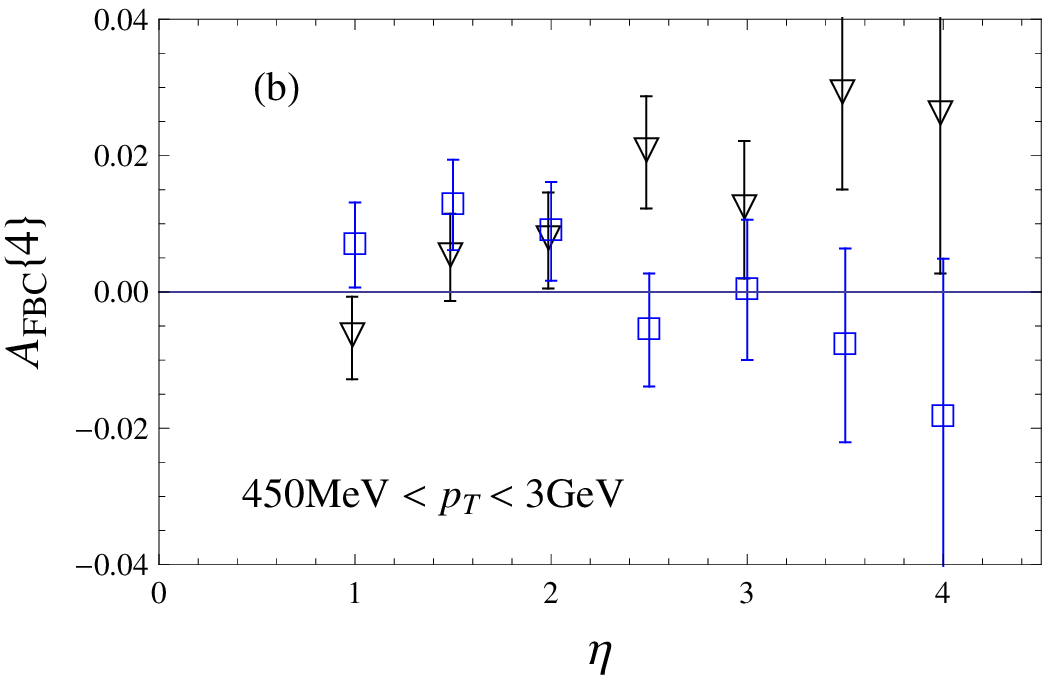}}
\caption{Pseudo-rapidity dependence of 2-particle (top row) and 4-particle cumulant using forward-backward rapidity bins (middle row) and using three rapidities (bottom row). In the left panels only primordial particles are considered whereas in the ones on the right resonance decays are included. In the latter case only charged pions, kaons, protons and antiprotons are considered and a cutoff in transverse momentum is included ($450~\mathrm{MeV}<p_T<3~\mathrm{GeV}$).}
\end{center}
\end{figure}

We propose experimental measures, which should be possible to apply with the available RHIC data, based on the use of cumulants involving particles from different rapidity regions.  If confirmed, this would support the initial assumptions concerning the fluctuations in the early stage of the fireball formation and asymmetric rapidity shape of the emission functions of the moving sources in the nucleus-nucleus collisions. Two- and four-particle cumulants using particles from forward and backward rapidity bins could be used for this purpose:
\begin{small}
\begin{align}
\cos(k\Delta_{FB})\left \{ 2 \right \} &\equiv \frac{\left \langle e^{i k(\phi_F-\phi_B)} \right \rangle }
{\sqrt{ \left \langle e^{i k(\phi_{F,1}-\phi_{F,2})} \right \rangle \left \langle e^{i k(\phi_{B,1}-\phi_{B,2})} \right \rangle} }
%\nonumber\\&
= \left \langle \cos (k \Delta_{FB}) \right \rangle_{\rm events}  +{\rm nflow} \nonumber\\
%\end{align}
%\end{small}
%\begin{small}
%\begin{align}
\cos(2k \Delta_{FB})\left \{ 4 \right \} 
& \equiv \frac{\langle  e^{i k [(\phi_{F,1}+\phi_{F,2})-(\phi_{B,1}+\phi_{B,2})]} \rangle}
  {\langle  e^{i k [(\phi_{F,1}-\phi_{F,2})-(\phi_{B,1}-\phi_{B,2})]} \rangle}
	=\left \langle \cos (2 k \Delta_{FB}) \right \rangle_{\rm events}  +{\rm nflow}.
\end{align}
\end{small}
Another possibility would be to use particles from three different rapidity bins (forward, central and backward):
\begin{small}
\begin{align}
A_{FBC}\{4\}&=
\frac{\langle  e^{i 2 [(\phi_{F}-\phi_{C,1})-(\phi_{B}-\phi_{C,2})]} \rangle - 
\langle  e^{i 2 [(\phi_{F}-\phi_{C,1})+ (\phi_{B}-\phi_{C,2})]} \rangle}
{v_{2,F} v_{2,B} v^2_{2,C}} \nonumber\\
&= \langle 2 \sin(2 \Delta_{FC}) \sin(2 \Delta_{BC}) \rangle_{\rm events} +{\rm nflow} 
%\nonumber\\
\approx
% &
 8 cov\left(\Delta_{FC}\Delta_{BC}\right)+{\rm nflow}.
\end{align}
\end{small}
The results of the {\tt THERMINATOR} simulation can be seen in Fig. \ref{cumulants} and suggest that it should be possible to detect the effect of the  torque. Full lines correspond to the expected result (without non-flow).
\section{Conclusions}
We have proposed experimental measures that should be able to detect a torqued fireball effect resulting from the fluctuation of the sources transverse position and their asymmetric emission profile. Due to the decrease in statistical noise with higher statistics and flow coefficient this effect should be looked for in mid-central or mid-peripheral collisions. Since the effect increases with $\eta$, the experimental search at RHIC should involve the forward and backward TPCs.
\section*{Acknowledgments}
Supported by Polish Ministry of Science and Higher Education, grants N~N202~263438 and N~N202~249235, and by 
the Portuguese Funda\c{c}\~{a}o para a Ci\^{e}ncia e Tecnologia, FEDER, OE, grant SFRH/BPD/63070/2009, CERN/FP/116334/201.

\bibliography{sqm2011_jmoreira}
\end{document}